\documentclass[compsoc]{IEEEtran}

\usepackage[utf8]{inputenc}
\usepackage[ngerman, english]{babel}
\usepackage{babelbib}
\usepackage{graphicx}
\usepackage{textcomp}
\usepackage{url}
\graphicspath{ {images/} }
\usepackage{xcolor}

\newcommand{\thething}{{WebEye}}

\title{\thething -- Automated Collection of Malicious HTTP Traffic}

\author{Johann Vierthaler, Roman Kruszelnicki, Julian Schütte\\
\textit{Fraunhofer Institute AISEC, Garching, Germany}}
\date{\today}

\begin{document}

\maketitle

\begin{abstract}
With malware detection techniques increasingly adopting machine learning approaches, the creation of precise training sets becomes more and more important. Large data sets of realistic web traffic, correctly classified as benign or malicious are needed, not only to train classic and deep learning algorithms, but also to serve as evaluation benchmarks for existing malware detection products.
Interestingly, despite the vast number and versatility of threats a user may encounter when browsing the web, actual malicious content is often hard to come by, since prerequisites such as browser and operating system type and version must be met in order to receive the payload from a malware distributing server. In combination with privacy constraints on data sets of actual user traffic, it is difficult for researchers and product developers to evaluate anti-malware solutions against large-scale data sets of realistic web traffic. In this paper we present \thething, a framework that autonomously creates realistic HTTP traffic, enriches recorded traffic with additional information, and classifies records as malicious or benign, using different classifiers. We are using \thething\ to collect malicious HTML and JavaScript and show how datasets created with \thething\ can be used to train machine learning based malware detection algorithms. We regard \thething\ and the data sets it creates as a tool for researchers and product developers to evaluate and improve their AI-based anti-malware solutions against large-scale benchmarks.
\end{abstract}

\section{Introduction}

The web is not only the largest pool of knowledge, entertainment, and cultural exchange of the planet, but also one of the most versatile attack vectors and in many cases one of the few unfiltered communication channels between enterprise networks and the Internet. With the advent of HTML 5 technologies such as local storage or web sockets, the client side plays an increasing role in modern web applications. Work load and logic is shifted from the web server to the browser and in many cases, the server side is reduced to a REST API while model, view, and controller of the actual application are delivered as JavaScript code to the client.

From the security perspective, the advent of web frameworks following this design pattern (e.g., Angular, Moustache, etc.) is favorable in general, as it reduces the attack surface at the server side. However, with browsers becoming more capable and JavaScript code bases executed by them larger and more complex, attacks in the opposite direction become more attractive: as HTTP and especially HTTPS is one of the main direct communication channels between clients in internal enterprise networks and servers in the public Internet, it becomes an attractive vector to deploy attacks against the client or even to use the client to launch attacks against enterprise networks. Examples are JavaScript-based port scanners, cross-site request forgery (CSRF) attacks or attempts to uniquely identify users across browsers and devices~\cite{COOKIEMONSTER}. Attack frameworks like BeEF~\cite{BEEF} illustrate the almost unlimited possibilities available to an attacker, once the client's web browser is "hooked".

One approach to counteract such attacks to the client side launched by malicious web servers is to inspect and classify traffic either in trusted proxies or directly in the browser. While traditional approaches attempt to detect pattern-based signatures in traffic, classic machine learning (ML) and deep learning techniques play an increasing role in classifying traffic to detect obfuscated or polymorphic malware and previously unknown attack vectors. To develop and evaluate such ML based attack detection techniques, it is mandatory to have large data sets of realistic web traffic with correct labeling of benign and malicious traffic. Without such data sets, vendor claims with respect to the quality of their malware detection solutions cannot be assessed and compared in an unbiased way and developers of ML based attack detection tools have no benchmark to evaluate their algorithm against. This is of special importance for deep learning approaches, where the quality of a neural network cannot be adequately assessed a priori.

However, the generation of high quality web traffic data sets is not trivial. First, actual traffic from real users is hardly available with the exception of few data sets (e.g., \cite{CTU13}) which are however extremely limited and biased, as users need to consent to take part in a traffic collection field test. Collecting web traffic from real users without their consent is obviously not possible or desirable. The browser history and form data submitted to web pages is regarded as personal identifiable information (PII) and allows to create detailed user profiles. For good reasons, the collection of such information without the user's consent and without anonymization is prohibited in most jurisdictions. Thus, the most promising approach is to generate realistic traffic by simulating the behavior of individual users following links and logging into web portals.

Another challenge is the labeling and feature extraction of collected web traffic. To create training sets for machine learning algorithms, data must be labeled as benign or malicious a priori.  While most ML algorithms tolerate errors in the training set, any mislabeling decreases precision. It is thus necessary to collect and aggregate "verdicts" from different sources for each accessed page and to extract features which are relevant for a training set but do not lead to overfitting.

In this paper, we address these challenges by introducing \thething{}, a framework to generate web traffic data sets for training malware detection solutions. While \thething\ is designed to create datasets, it can also be used in a continuous learning mode for adaptive anomaly detection systems. We introduce the main components of \thething{} and illustrate how ML based algorithms can be integrated for the creation of initial labeling verdicts, as well as their accuracy can be assessed a posteriori by evaluating against the created data set as a benchmark.

In section \ref{sec:related-work}, we review work related to ours that influenced the creation of the framework and discuss how it differs from our approach.
Section \ref{sec:traffic_collection} introduces our overall approach and its capabilities and limitations.
The next section outlines the overall technical setup that was used for creating and testing this framework and gives detailed descriptions of the individual components the framework consists of.
Section 4 explains how the framework was used for the creation of a dataset and the specific setup, performance, experiences and steps for accessing the dataset are discussed there.
Finally, the conclusion generally summarizes the framework and its capabilities, but also its limitations.
Furthermore an outlook on the potential fields of applications for this frameworks is given.

\section{Related Work}
\label{sec:related-work}
  Work related to ours in concerned with the creation of datasets of web traffic suited for training and evaluation of malware detection solutions. For obvious reasons, the amount of publicly available datasets of real-world user browsing behavior is scarce. Justified data privacy and secrecy concerns forbid capturing and publishing traffic of enterprises or even of major Internet nodes.

  Most similar to our work is the HTTP Dataset CSIC \cite{CSIC2010}, an automatically generated dataset of web traffic, including normal and malicious traffic. Similar to us, the authors intent to use this dataset for training and evaluation of attack detection mechanisms, but there are also notable differences: first, the CSIC dataset includes only requests to a single (artificial) e-commerce application and thus only serves as a training set for detecting client-side attacks against similar types of web applications. We focus in contrast on the detection of server-side attacks against the client and thus need to include various web applications in the wild and the content they deliver to a real-life user. Second, CSIC is a pure dataset, but not a framework to generate such sets. The generation methodology is not precisely defined and so updating the CSIC is impossible. That fact that it is from 2010 makes it unsuited for training of newer HTTP 5 based attacks. Finally, with 36,000 normal and 25,000 malicious requests, the CSIC set is significantly smaller than training sets typically required for training deep neuronal networks.

  CTU-13 \cite{CTU13} is a 1.9 GB dataset of benign and botnet traffic that was captured and manually labeled in the CTU University, Czech Republic, in 2011. The dataset itself differs from ours in that it has been collected passively across all layers and does not reflect individual browsing behavior. The methodology of manually labeling records is tedious and thus potentially unreliable and does not scale to the amount of data that we set out to create.

  Different from our work are also pure malware data sets referring to actual malware samples such as executable files and their behavior. Such data sets are widely available, such as the BIG 2015 data set hosted by Microsoft which contains 500 GB malware samples \cite{BIG2015}, the Kharon dataset of reverse engineered Android applications \cite{kharon2016}, Drebin -- another Android malware dataset --\cite{drebin2013,drebin2014}, or malware sample behavior such as the dataset of malware system calls in \cite{trinius2009}.

\subsection{Collection and Analysis}
  Canali et. al proposed with \textit{Prophiler}~\cite{PROPHILER} a machine learning based filter for static upfront classification of malicious web pages in order to collect and verify actual malware in a more efficient manner. In order to train this classifier, various, e.g., HTML and JavaScript based features are extracted from benign and verified malicious contents.  

\subsection{Zozzle}
	Microsoft presented with \textit{Zozzle}~\cite{ZOZZLE} a work, that focuses on machine learning based detection of JavaScript malware. 
	The approach proposed in \textit{Zozzle} is a Bayesian classifier, which operates on ASTs generated from the JavaScript source code. 
	The resulting abstract syntax tree is then analyzed if it consists of subtrees or leafs, that indicate the presence of malware. 
	In this work we also form an AST out of given JavaScript code and use it as an approach for simplifying feature extraction. 
	However, our approach differs from Zozzle, as we also include various HTML features typically involved in malware when setting up the feature vector.
	This allows us to detect malware, that does not utilize JavaScript in its attacks.
	Additionally, instead of a Bayesian classifier, we are using a RandomForest-Classifier, as it delivered the overall best performance when compared to other classifiers.


\section{Traffic Collection Framework}
\label{sec:traffic_collection}
  The framework was developed with the primary intention of simulating realistic website browsing behavior with automated user agents in order to gather malicious HTML and JavaScript files in a way that allows us to trace the various stages of actual attacks on clients, e.g., bootstrapping and exploitation.
  We classify the resulting contents with the help of signature based detection methods (ClamAV), blacklists (Google Safebrowsing) and machine learning. 


\subsection{Overall Concept/Architecture of \thething}
\label{subsec:tc_concept}
The abstract concept of the framework consists of 5 basic components: 

\begin{itemize}
	
	\item \textbf{Capturing Component/Sensor:} The capturing component must be able to record the traffic of one or more user agents at the same time without severely sacrificing responsiveness.
	After the full data for a traffic flow has been acquired, the data record is then forwarded to the next stage for further processing.
	
	\item \textbf{Automated User Agent:} The user agent automatically interacts with different websites to generate HTTP traffic to collect. 
	Besides the generation of data, the user agent should also be able to mimic the browsing behavior of a real user -- to some extent -- to create a more realistic traffic. For example by attempting logins to web services, scrolling the document and interacting with buttons and other HTML UI elements.
	
    \item \textbf{Data Augmentation Facilities:}
    Collecting and deriving additional data provides us with a broader knowledge about the subject. 
    This knowledge can, for example, be used in the development of additional detectors, e.g., machine learning classifiers, since the derived data may be used as a supporting feature-set.
  
	
  \item \textbf{Malware Detection Facilities:} Detectors represent the upfront classifiers, that are used to determine if a HTTP-traffic flow carries malicious content. 
  They should fulfill the following requirements:

  \begin{itemize}
    \item \textbf{Fast Processing:} The detectors have to quickly provide their verdicts in order to ensure responsiveness.
    \item \textbf{Updatability:} For reliable detection, the classifiers should be up to date or should provide the necessary update-functionalities.
  \end{itemize}

  \item \textbf{Storage Component:}
  The storage component has to fulfill two requirements:
  \begin{itemize}
    \item The database should be able to contain non homogeneous records, e.g., datasets of variable structure.
    This means, that the storage unit should also be able to provide a lookup functionality, which allows us to query records for fields, which are not present in other records.
    \item Storing/retrieving files and linking them to a data record should be easily possible.
  \end{itemize}
  
  	\item \textbf{User Agent Management:} In order to have a good control over the number and type of user agents active in the system, a management component is used.
  	This component should be able to carry out the steps necessary for creating, monitoring and removing the active user agents.\\
  
\end{itemize}

\begin{figure}[tb]
	\centering
	\includegraphics[width=\columnwidth]{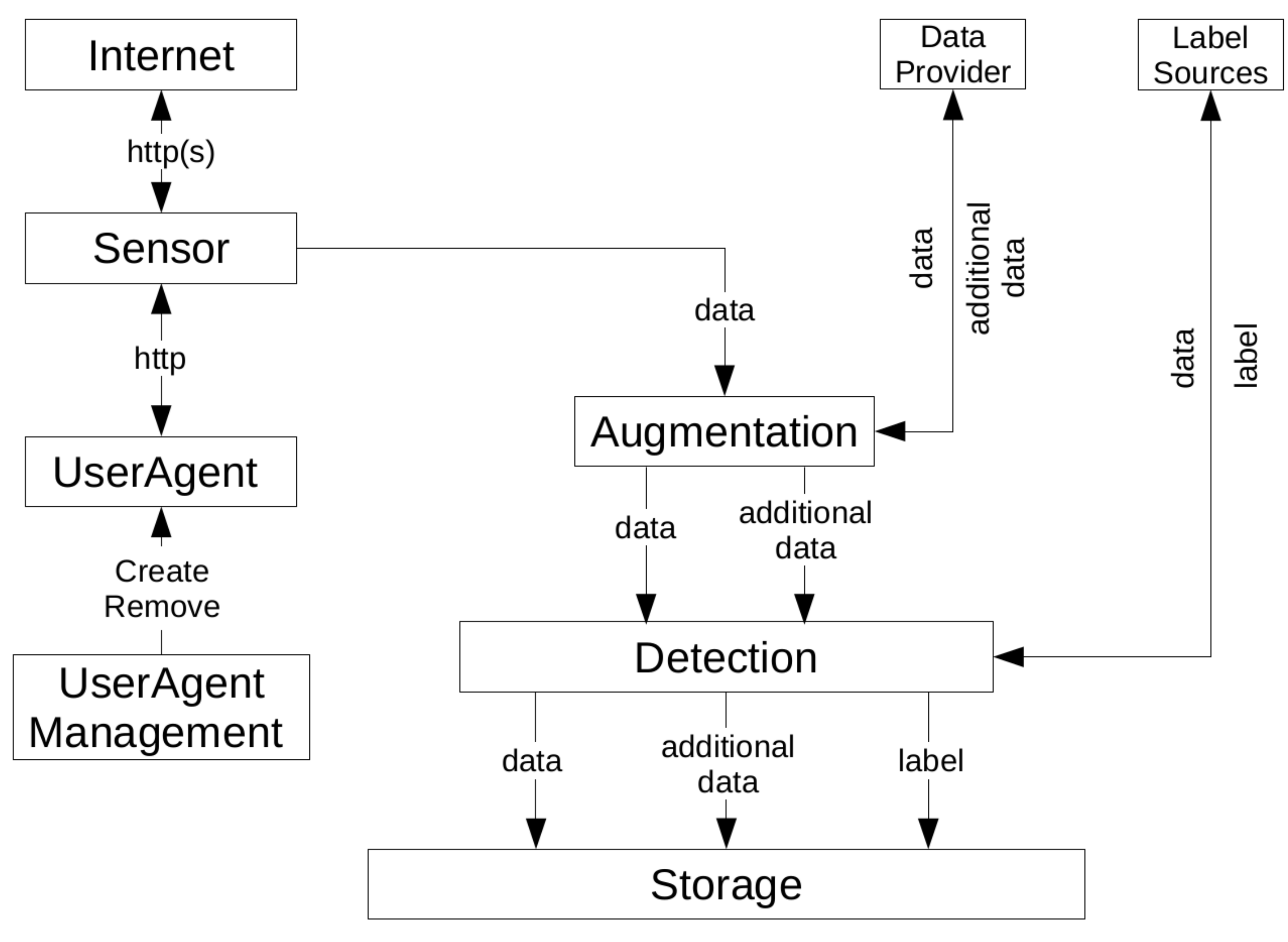}
	\caption{Conceptual composition of \thething }
	\label{fig:concept}
\end{figure}

\newpage
\subsection{Implementation and Setup}
\label{subsec:tc_implementation}
  This section provides a detailed description of the utilized technologies and implementation.
  It should be noted, that we are using virtual machines within an Openstack private cloud environment to realize the components of \thething.

  \subsubsection{Implementation: Capturing Component/Sensor}
  \label{subsubsec:icap_squid}
    The capturing component consists of both, a ICAP-Server for handling the data and a Squid Proxy Server, which transports the data from the user agents to the ICAP-Server and vice-versa. 

    \paragraph{ICAP-Server}
    \label{par:tc_icap}

    An Internet Content Adaption Protocol Server (ICAP-Server) component is the primary component of our framework. 
    This ICAP server is contacted by the Squid Proxy for every HTTP request and response initiated by the active user agents. 
    It is primarily responsible for recording, enriching and storing data resulting from HTTP requests, responses and the corresponding bodies. 
    To enrich the data the ICAP server queries the implemented detectors for malware determination and augments the existing data with the help of the employed data enrichment facilities.
    The requests, responses, response-bodies, verdicts and data augmentations are then stored in the data storage component.
    Due to the characteristic of the ICAP protocol, this component can also work as an enforcement module, warning users about malicious content.
    Another reason for choosing ICAP is the fact, that it is a standard way for HTTP filtering, what simplifies deployment in existing networks.

    \paragraph{Squid Proxy Server}
    \label{par:tc_squid}
    The Squid proxy is the sole client to the ICAP server within our framework. 
    It considerably simplifies HTTP traffic recordings, since the Squid proxy is capable of the ICAP protocol and the client components~(~\ref{subsubsec:client_component}) only have to be configured to communicate via the Squid instance. 
    The proxy can also be configured to act as a Man-in-the-Middle, which opens secured connections (so called SSL Bump) and hence enables us to also analyze and store otherwise encrypted traffic.

  \subsubsection{Implementation: Automated User Agent}
  \label{subsubsec:client_component}
    This section describes the component of the framework, which acts as a client towards a targeted website.
    It elaborates on the utilized technologies, sources of URLs and implemented website interaction strategy.

    \paragraph{Browser Automation with Selenium}
    \label{par:selenium}
      One of the earliest decisions in the development of the framework concerned the user agent, which is responsible for communicating with the targeted websites.
      The decision was against a static crawler, since many malware distributing websites utilize evasive strategies and methods in order to remain undetected by anti-malware companies and technologies~\cite{REVOLVER}.
      For example, a bootstrapping website sets the cookie via JavaScript to a specific value and only with this cookie, the next request against the server will deliver the malicious payload.
      Such evasive countermeasures nullifies the reasons to use (faster) static crawlers.
      Therefore, we used the browser automation framework \textit{Selenium} to deploy and control popular browsers (e.g., Chromium and Firefox).
      This browser automation not only allows malware bootstrapping procedures to take place, but also enables us to easily simulate a basic user-website-interaction. 
      Despite promising a substantial boost in performance compared to Firefox or Chromium, we decided against headless browsing, e.g., PhantomJS~\cite{PHANTOMJS}, due to them being discernible from popular browsers, when fingerprinting it for certain objects, e.g., WebGL-Components.
      
      The employment of the Selenium framework makes our system browser-agnostic and allows for using any browser, as long as it implements the WebDriver API, which most popular browsers do. This also enables us to quickly integrate new browsers in the future.

    \v{Website Interaction Strategy}
    \label{par:website_interaction}
      To implement an approximately realistic user simulation, a certain website interaction strategy has to be considered. 
      Such a strategy enables us to collect data more efficiently and also reduces the risks of being deflected by evasive mechanisms of more elaborate malware. 
      One of the first considerations was to set the screen size for the browsing window to a common value (e.g., 1366 x 768), this contributes to the goal for letting the user agent appear more like an actual person using a browser. 
      Another important decision was the integration of \texttt{Flash} into the browsers. 
      This enables us to record \texttt{Flash} based exploits.

      For the actual interaction itself, the selenium logic "opens" a seed URL, waits for the page to load and then scans the website for a possibility to log into the service. 
      If such a possibility is found, the user simulation queries the web service \textit{BugMeNot}~\cite{BUGMENOT} for publicly shared account credentials for the specific website and then proceeds to fill out the login form and sends it to the server.
      In case no credentials are available, the selenium logic uses predefined values. After a login attempt, the selenium script filters the DOM for elements to interact with (e.g., clicking on links, buttons and other inputs). 
      This interaction continues until a configured number of interactions is reached or the pipeline of elements to interact with is depleted. 

    \paragraph{URL Seeders}
    \label{par:seeders}
      As the user agent relies on URLs to start crawling a website, a large source of URLs has to be provided.
      A vast quantity of malicious or benign URLs is important to build a reliable training set. 
      For the framework we utilized so called seeders that prepare sources of domains, IP addresses and URLs for the user agents. 
      Each deployed user agent uses only one seeder.

      For this paper we utilized 3 seeders, for which each had a different focus for finding contents:
      \begin{itemize}
        \item \textbf{Alexa 1 Million:} The first seeder of the framework provides a large quantity of URLs for largely benign websites. 
        Alexa 1 Million is a website ranking which lists the top 1 million most popular websites. 
        This ranking of (mostly benign) websites results in a very low probability of finding malicious contents. 
        This is important, since reliable predictive systems also need to process benign contents correctly.

        \item \textbf{MalwareDomainList:} Due to the low probability of finding malware with the Alexa 1 Million seeder, an additional seeder was implemented, which had the focus on malware related URLs.
        This seeder uses the URLs provided by the web service \texttt{MalwareDomainList.com} to deliver possibly malicious URLs to the user agents.

        \item \textbf{Openphish:} This seeder specializes on phishing websites, which may contain phishing schemes as well as malicious content. 
        This seeder primarily complements the other seeders by providing URLs to supposedly malicious and/or compromised servers.
      \end{itemize}

      Using specialized seeders instead of arbitrarily dialing URLs considerably improved the probability of finding malicious contents and with additional, different seeders more malware could be collected within shorter time.

    \paragraph{Monitoring and Maintenance}
    \label{par:manual_monitoring}
      It is important to (visually) monitor the behavior of the client component, as it not only displays certain fault cases, but also enables us to observe how malicious websites impact the browser of our user agents. 
      This observation is realized by using a display via a XVNC backend, which enables us to (visually) observe the browser window and the rendered contents within it. 
      Since the user agents are isolated in a fortified subnet and a direct connection was not intended, we are using port forwarding on the squid proxy machine to establish this connection. 
      SSH access for maintenance/testing was also implemented with port forwarding.

  \subsubsection{Implementation: Data Augmentation Facilities}
  \label{sec:data_enrichment}
    In order to provide informative statistics about the collected data, we integrated so called data augmentation components, which fetch additional information from external sources or derive data or meta-data from the analyzed datasets.
    This enables us to find correlations between datasets and may be used to strengthen the indications if a content is malicious or not.

    \paragraph{GeoIP}
    \label{par:geoip}
      GeoIP enables us to determine meta-data, e.g., the approximate location of a server by the server's host name or IP address. 
      With the city level GeoIP database this approximation is more detailed and allows us to make statements, for example, about specific malware related regions.

    \paragraph{WhoIs}
    \label{par:whois}
      WhoIs provides meta-data about, amongst others, the time of validity of a specific domain. 
      Malware domains are often registered for a rather short time and therefore WhoIS can provide indications for malware predictions.

  \paragraph{HTML/JS Feature Extraction}
  \label{par:feature_extraction}
      In total we are using 58 different features, which are used for determining the maliciousness of a presented content.
      These features are based on the \textbf{Prophiler}~\cite{PROPHILER} paper as well as manual evaluation of suspicious files.
      In the following list, five example features are explained (all features are listed in Table~\ref{extracted_features}): 
      
      \begin{itemize}
        \item \textbf{Number of script strings in HTML or JavaScript files:} DOMs of malware websites are often cluttered with JavaScript-nodes (script-tags in the raw HTML) or malicious JavaScript scripts inject additional script nodes into the DOM to carry out an attack. 

        \item \textbf{Shell code probability in strings:} The probability itself is based on the entropy each string in a script possesses.
        The higher the entropy of a string is, the higher the probability, that a string contains shell code-like information.
        Shell codes are often used in, e.g., heap spraying attack, which allow the attacker to carry out harmful actions on the clients computer. 
        But the shell code probability might also indicate the utilization of obfuscation or encoded images, since blobs of characters might also resemble shell code.

        \item \textbf{Number of \texttt{eval} strings:} The JavaScript function \texttt{eval} receives and executes a string of JavaScript code.
        It should be noted that the use of eval is by itself not malicious, however, eval is a common pattern in malware used, for instance for obfuscation or for executing asynchronously loaded strings containing JavaScript code.

        \item \textbf{Number of \texttt{Iframe} strings:} Malware websites have a tendency to utilize one or more Iframes to carry out, e.g., drive-by-attacks.
        Iframes are, similar to eval, a common theme in malicious contents.

        \item \textbf{Strings-to-Script-Ratio:} Malware often tries to evade signature based scanners by, e.g., embedding one or more strings, which contain a harmful payload in an obfuscated format, in a, for example, JavaScript file.
        During runtime, this string is deobfuscated and executed, e.g., with \texttt{eval}.
        Depending on the payload, this/these string/strings may represent the majority content of the script. 


      \end{itemize} 

      Storing these features enables us to develop malware detectors, which operate on this information and allows more possibilities for data analysis.
      Additionally, these features can be used to have an approximation on the design of nowadays HTML/JavaScript malware.

  \subsubsection{Implementation: Malware Detection Facilities}
  \label{subsubsec:oracles}
  This section elaborates on the detection components utilized in this framework.
  These detectors are responsible for detection and prediction of potentially harmful traffic. 
  For this framework, we rely on three different label sources to determine if a certain record is malicious or not. 
  These labels can then be used to develop malware detection tools, e.g., machine learning based classifiers.

    \paragraph{Label-Source: Google Safebrowsing}
    \label{par:safebrowsing}
      Google offers a service to determine if the resource of a URL is deemed malicious or not. 
      Google Safebrowsing hereby matches the URL -- more specifically a hash prefix of an URL -- against a blacklist maintained by Google. 
      A Safebrowsing-query may result in five different results. 
      These types are:
      \begin{itemize}
        \item Malware
        \item Social Engineering (e.g., phishing)
        \item Unwanted Software
        \item Potentially Harmful Applications
        \item Unspecified
      \end{itemize}


    \paragraph{Label-Source: ClamAV}
    \label{par:clamav}
      ClamAV is a signature based anti virus engine, which we incorporated into our collection framework to leverage an additional detector to label malicious content. 
      In order to keep the ClamAV signatures up to date, we are utilizing the ClamAV signature update service \textit{Freshclam} in regular intervals. 
      We are using the network socket of the ClamAV daemon \textit{clamd} for file scans and the subsequent report retrieval. 
      Since the framework stores the response bodies in the format the server delivered it (compressed or uncompressed) to us, we are decompressing the files (based on the headers in response) before sending it to ClamAV.
      This helps us to avoid false-negatives originating from ClamAV not being able to fully evaluate certain compressed formats or chained compressions. 

    \paragraph{Label-Source: Virustotal}
    \label{par:virustotal}
      Virustotal is a web service for scanning URLs or files for malicious content by using multiple anti virus engines. 
      The main strength of Virustotal is the utilization of more than 50 anti virus engines to analyze a single file, which enables us to automate more accurate malware predictions. 
      However, the slow scanning speed and the public API request restrictions (4 requests per minute) of Virustotal makes synchronous report retrieval impractical for our use case. 
      Therefore, the Virustotal component only scans files, that were declared as malware by at least one of the other detectors. 
      Also, the scanning itself follows an asynchronous approach. 
      This approach utilizes 4 status flags for a Virustotal report (unscanned, scan in progress, scan finished and error), which are either set by the ICAP component or the Virustotal component.
      For each request the ICAP sets the status flag of the report to \textit{unscanned}.

      Concurrently, the Virustotal component uses two threads:
      \begin{itemize}
        \item One thread queries the database for unscanned files and proceeds to decompress the files (by using the algorithms specified in the response headers) and subsequently uploads those files to the Virustotal service.
        After the upload, the status flag is set to \textit{scan in progress} and the Virustotal scan id is added to the Virustotal field in the database.
        Any error changes the flag to \textit{error} (e.g., request exceeds the file size limit of Virustotal).

        \item The other thread queries the database for files with the \textit{scan in progress} flag and retrieves the corresponding report with the scan id.
        If a report could be retrieved, the record is written into the database and the status flag is set to \textit{scan finished}.
      \end{itemize}

	\subsubsection{Implementation: Storage Component}
	\label{subsubsec:tc_datastorage}
	\paragraph{MongoDB and GridFS}
	\label{par:tc_mongodb}
	For storing the dataset a cluster of MongoDB shards is used that represents the storage behind \thething. 
	It is responsible for storing data of requests, responses, content, labels and other additional data. 
	Our current setup consists of five MongoDB shards, but this can be scaled upwards or downwards, if needed. 
	A NoSQL Database was chosen because it offers a very flexible way to store data that does not necessarily follow a common schema, i.e., are not homogeneous in structure.
	For example, the individual headers within the HTTP-Requests and HTTP-Responses may vary in different fields, e.g. header-fields for content-type or content-encoding.
	Data of requests, responses, etc. are fitted into structured documents, that are then stored into the normal MongoDB database while the content delivered by the response is saved separately into GridFS. 
	This separation is needed as the content may vary heavily in size and may exceed the BSON-document size limit of 16 MB~\cite{MONGODB_GRIDFS}.
	In order to avoid duplicate files in the GridFS, a SHA-1-Hash of the body is calculated.

	\paragraph{Data Exploration with Elasticsearch and Kibana}
	\label{par:tc_elastic_search}
	During the evaluation of past datasets, we decided that effective data exploration and visualization is not easily achievable with MongoDB alone.
	Therefore, we incorporated Elasticsearch and Kibana for data exploration.
	This exploration is made possible by exporting the MongoDB datasets to Elasticsearch.

	\subsubsection{Implementation: User Agent Management}
	\label{subsubsec:tc_client_mgmnt_component}
	In order to control the user agents active in \thething, a central user agent management entity was required that is able to create, tear down, monitor and reboot user agents. 
	The primary use cases for this management component are the creation of different types of user agents and the restart of user agents in case of lockups, e.g., due to websites opening browser native dialogs (e.g., printing dialogs, alerts and basic-authentication), slow loading times or browser restart problems.
	This was realized by implementing a REST client, which interacts with the API provided by the OpenStack cloud controller.
	This API enables us to create, remove and reboot user agents instances.
	In order to find and subsequently reboot locked up user agents, we regularly query the MongoDB database if the last record of a user agent is older than a certain threshold, e.g., 10 minutes.

  \subsubsection{Securing and Scaling the Deployment}
  \label{subsubsec:dep_openstack}

      \paragraph{Security}
      Since \thething\ uses actual browser engines to collect web traffic, the risk of compromise through exploits has to be considered and appropriate measures of fortification have to be implemented in order to protect the framework and the platform it is built upon. The main attack vectors we need to protect against are attacks against the internal network of the framework, e.g. by Denial of Service (DoS) attacks launched in the client browsers and infections of either framework-internal machines or even external endpoints through executable code. Information stealing attacks such as history sniffing or cookie stealing are less relevant, as \thething\ does not use any secret login data or tokens. 

      \thething\ limits the attack surface by deactivating any code execution except JavaScript so that exploits over Silverlight or ActiveX cannot be launched against the framework. Further, user agents run in isolated unprivileged containers and have no means of interfering with the network, memory, or file system of other user agents. Each user agent is located in an isolated, virtualized subnet, which offers no direct connection to the Internet and other local components. The only allowed connection is the one with the HTTP(s) squid proxy server. This is enforced by an \texttt{iptables} firewall configuration on the squid machine that drops every connection which is not addressed to a port of the proxy server.
 
      As the exposed components (browsers) are isolated in an virtual environment, a potential exploitation should be contained. The only allowed communication is with open Internet. This could theoretically allow an attacker, who took over an UserAgent, to attack a website using the UserAgent as a proxy. This risk of collateral damage is mainly remediated by terminating the user agents and starting again from a fresh image, which is known to be benign, in regular intervals.

      \paragraph{Scalability}
      
	  It is important to be able to dynamically control the number and type (e.g., user agents with malware seeders) of user agents in the system.
	  This gives us a control over what kind of data and in which quantity/speed it should be collected by the framework.
	  To achieve this, we bundle each component into a docker container, which is then shipped to machines within a cluster.
	  Scaling the user agents is done by the user agent management component of \thething, which would invoke/remove user agent instances.
	  This allows us to dynamically deploy different flavors of user agents (e.g., a Firefox browser with a phishing focused seeder) in a simple manner.
	  The scalability is realized by utilizing Openstack's interfaces and appropriate storage facilities.
	  With this scalability functionality in place, the remaining bottleneck is the hardware and/or the available bandwidth.


    
\begin{figure}[tb]
	\centering
        \includegraphics[width=\columnwidth]{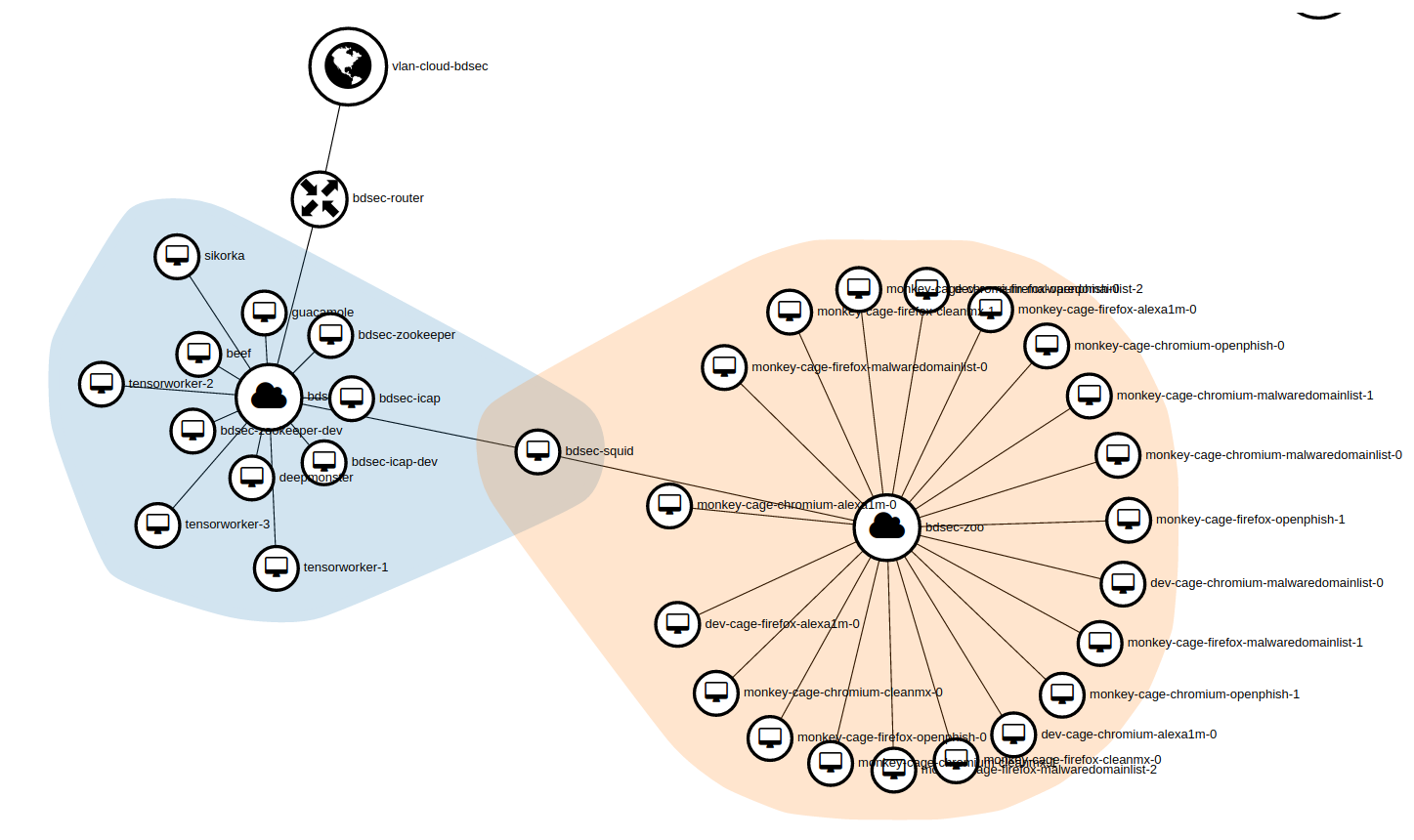}
        \caption{OpenStack Network}
	\label{fig:openstack}
\end{figure}


\section{Dataset Evaluation}
\label{sec:dataset_evaluation}
  This section of the document presents an evaluation of the traffic data, which was collected within 2 months of crawling. 
  The dataset was generated with user agents of two types of browsers (Firefox and Chromium) and different URL sources.
  For each type of browser one user agent for benign, two user agents for phishing focused and five user agents for malware focused samples were utilized.


    In total, the crawlers created 43 million request/response-pairs for later analysis.
    The unique count in this work is always based on a SHA-1 hash of the response body. 

\begin{figure}[tb]
	\centering
    \includegraphics[width=\columnwidth]{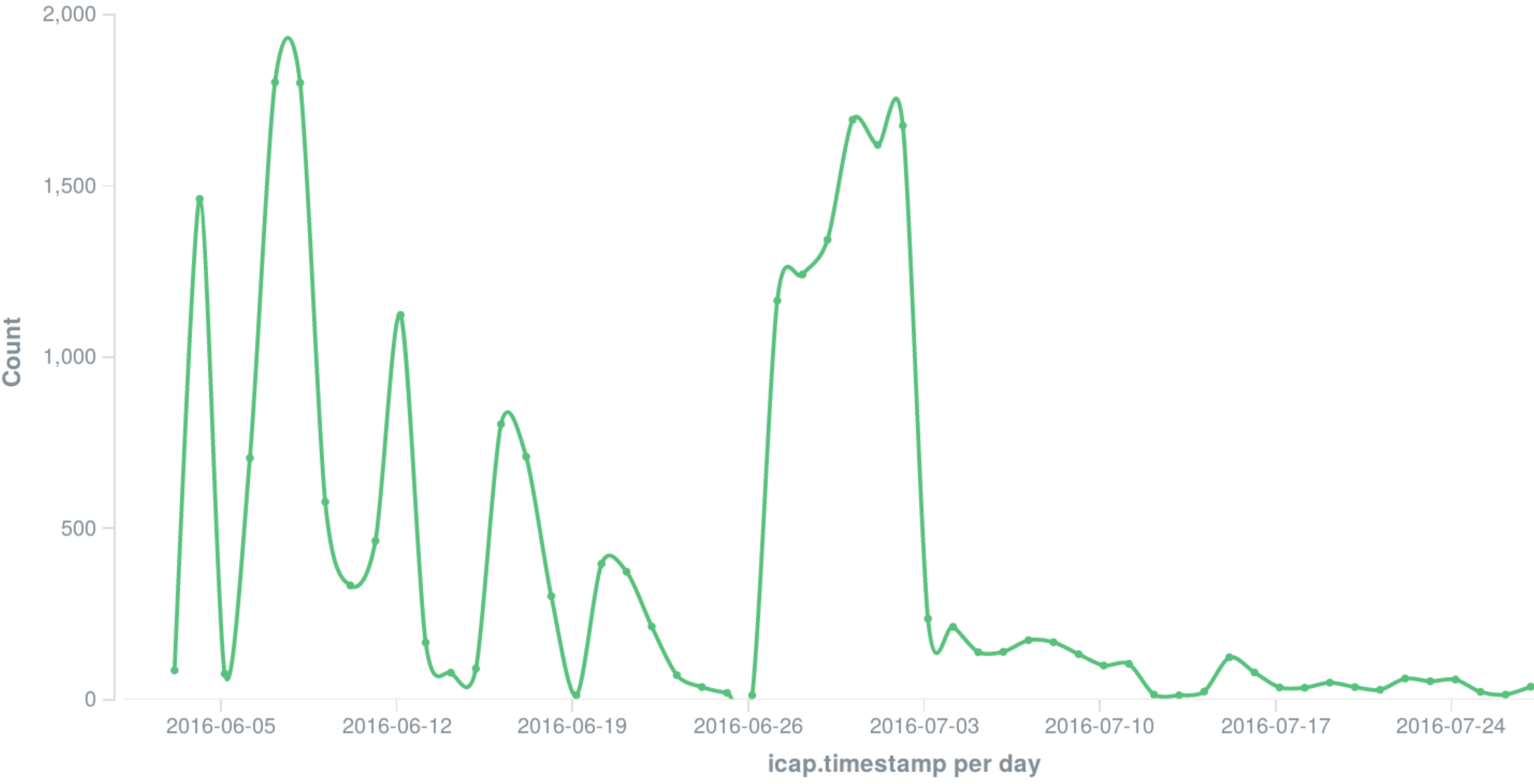}
    \caption{Progress of malware collection over time. }
	\label{fig:figure1}
\end{figure}

    The figure above shows the number of classified malware (checked against our ground truth) collected during the collection phase of \thething.

\begin{figure}[tb]
	\centering
        \includegraphics[width=\columnwidth]{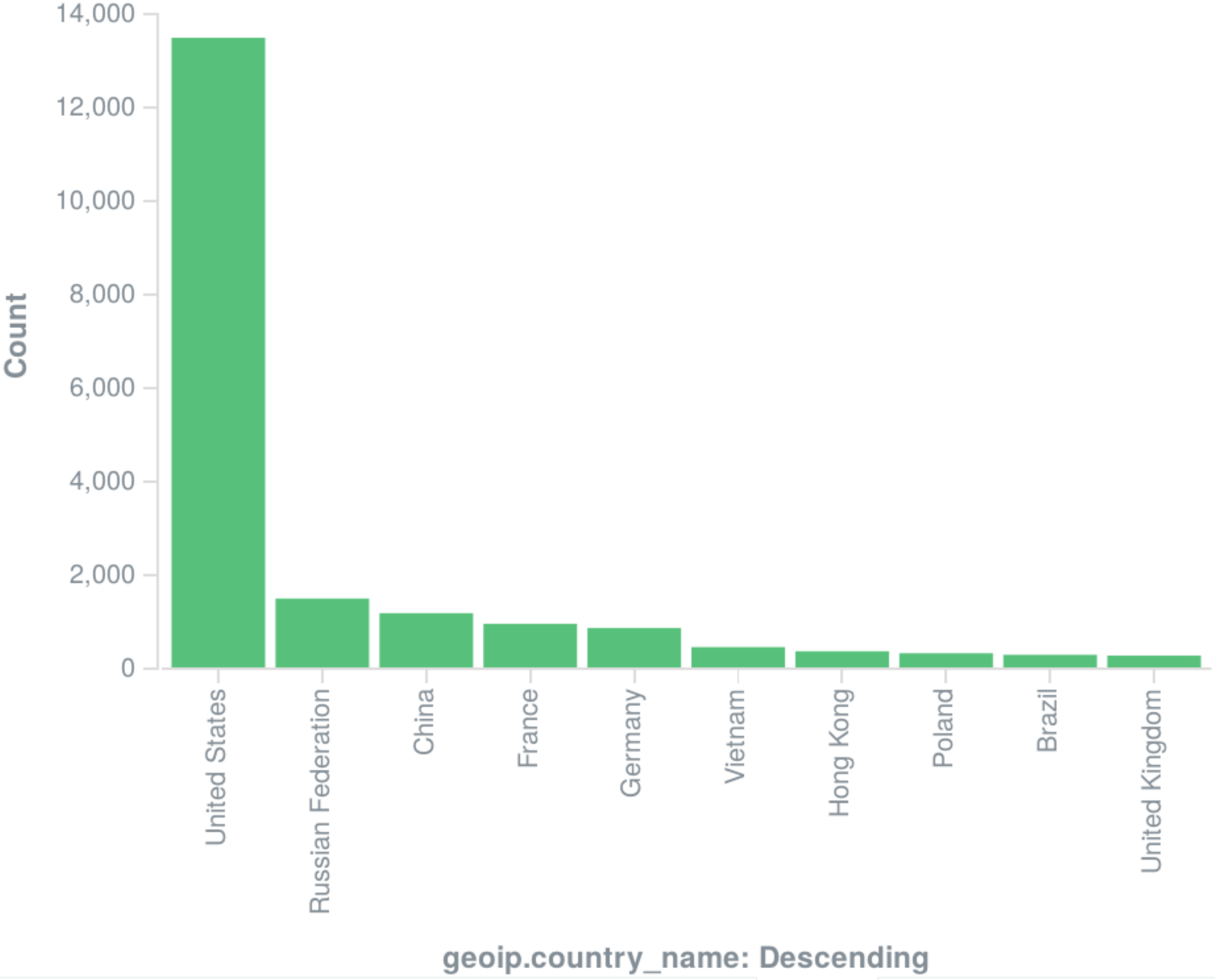}
        \caption{Top 10 malware countries}
    \end{figure}

    This figure presents the top 10 countries hosting malware. 
    The data about the location of the hosting server is based on GeoIP. 

	\begin{figure}[tb]
		\centering
        \includegraphics[width=\columnwidth]{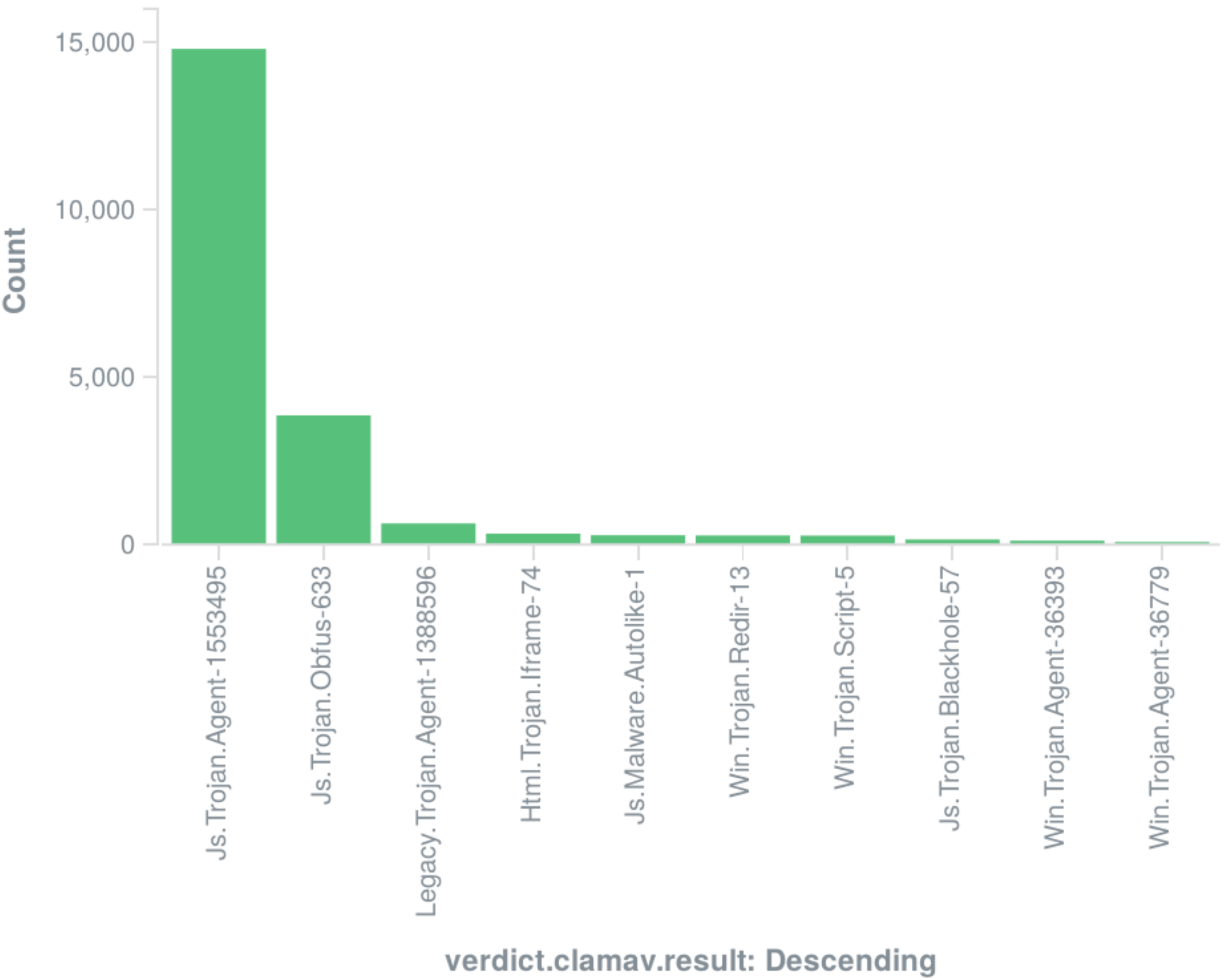}
        \caption{Top 10 malware types (based on clamav)}
  \end{figure}

	The diagram above shows a Top 10 of the most encountered types of malware based on the labels delivered by ClamAV.

  \subsection{Evaluating the label sources}
  \label{subsec:de_malware_prediction}
    The following statistics show the number of detections within the database and primarily makes comparisons to a malware ground truth to show the accuracy of an malware detector. 
    Due to the discontinuation of \textit{Wepawet}, a Virustotal scan with at least 10 alerts serves as our ground truth for malware in this work.


    \subsubsection{Google Safebrowsing}
    \label{subsubsec:de_safebrowsing}
      \begin{itemize}

        \item \textbf{MALWARE:} 104751 (unique) responses are stored within the database, that were flagged as malware by Google's Safebrowsing Service.
        However, if compared to the specified ground truth, the number drastically diminishes to only 10373 requests. The number of flagged responses with no Virustotal alarms is 93015.

        \item \textbf{UNWANTED\_SOFTWARE:} 9737 responses were flagged as unwanted software. Unwanted software does not necessarily indicate actual malware in the traditional sense. However, 306 out of those requests were deemed malicious by Virustotal.

        \item \textbf{SOCIAL\_ENGINEERING:} 62697 responses were flagged as websites, that attempt, e.g., phishing. Since social engineering is not considered malware, a Virustotal scan was not performed, except when ClamAV raised an alert.

        \item \textbf{POTENTIALLY\_HARMFUL\_APPLICATIONS:} No responses with this flag exists within the database.

        \item \textbf{THREATTYPE\_UNSPECIFIED:} No responses with this flag exists within the database.

      \end{itemize}

    \subsubsection{ClamAV}
    \label{subsubsec:de_clamav}
      ClamAV flagged 21651 requests in our database as malicious. Compared to our ground truth (at least 10 anti-virus-engines raised an alert), 21306 samples are actually malware. The delta consists of samples that have less than 10 detections.

      ClamAV detected the following signatures within the majority of the dataset:

      \begin{itemize}
        \item \textbf{Js.Trojan.Agent.1553495:} This signature was detected in 14807 distinct files of the dataset.
        \item \textbf{Js.Trojan.Obfus.633:} 3859 files matched to this signature.
        \item \textbf{Legacy.Trojan.Agent-1388596} concludes the Top 3 with 631 findings
      \end{itemize}



    \subsubsection{Virustotal}
    \label{subsubsec:de_virustotal}
      All in all we were able to scan 24928 malware samples, of which 22818 samples triggered at least 10 detections by Virustotal's antivirus-engines. The average detection count over the whole collection period is 26 detection.

      \subsubsection{Content-Types}
      In terms of content-types, the malware dataset primarily consists of the following types:
      \begin{itemize}
        \item{HTML:} The majority of the malicious dataset is represented by HTML files (21610), of which 21347 are HTML files with embedded JavaScript.
        \item{Octet-Streams (Binaries):} The second largest content-type falls far behind with a count of only 330 distinct files.
        \item{JavaScript:} With 329 files, JavaScript is the third largest party within the dataset. 
      \end{itemize}

      The remaining samples are distributed over different content-types, e.g., Android APKs, compressed files (rar, gzip, zip), images (jpeg, bmp), flash and plain text.

    \section{Exemplary HTML/JS-Malware Detection with Machine Learning}
    \label{sec:classifier_eval}
    
	 
	 

    As one of the main purposes of \thething{} is to enable researchers to develop and fine-tune ML-based detection algorithms, we demonstrate how \thething{} data sets can be used for training ML algorithms. For the sake of demonstration we choose a simple classic ML algorithm here, but in our research we also covered the development of deep learning approaches.
    
    Because machine learning approaches in malware detection (and deep learning especially) need large amount of labeled samples to learn reasonably, \thething{} enables researchers to collect data sets large enough to employ those algorithms for web user protection.
    
    In our case out of a 500 GB data set with 43 million samples, about 20,000 were labeled as malicious which indicates 0.5\textperthousand \ maliciousness ratio in the wild. At that ratio, existing data sets of only a few hundred mega bytes do not contain enough malicious samples to create reasonable training sets from. With about 20,000 distinct malware samples however, \thething{} was able to generate a sufficiently large number that allowed us to train a Random Forest classifier. Also, \thething{} allows to continuously update the data set to reflect the current malware landscape and allow re-training the ML systems to detect new threats. 
    The separation and asynchronous approach of the label collection process further allows updating labels even after the samples were initially collected. This allows for a posteriori improvement of the labels quality, because it gives AV companies time to find and identify new malware strains which can be considered in the a posteriori labeling, while \thething{} is working with real-time traffic data.


        As an example for any machine learning classifier to train with our data set we chose the random forest classifier~\cite{Breiman2001} implementation of the Scikit-Learn~\cite{SCIKIT}  library. The algorithm operates on the above-described HTML/JS features that were extracted from the collected JavaScript or HTML samples.
        
        The first step in training the classifier was the creation of the training and testing sets.
        For the training set of malicious samples we divided the collected malware dataset into approximately two halves (11861 training records vs. 10092 testing records). 
        For training benign data, we took the tenfold amount of samples, which did not raise an alert from any of the oracles. 
        In order to test the classifier's performance with benign data, we used a testing set of 10000 benign samples.
         
        Due to the imbalanced amount of malicious and benign samples, the classifier was trained with a 1:10 weighting.
        The classifier operates with a maximum of 10 trees in the forest, has no maximum depth limitation and uses the Gini impurity measure.

        As can be seen from the following tables, the classifier correctly detects 0.99\% of actual malware as such and thus has a high true positive rate, but would need further optimizations with respect to its false positive rate. 
        Using data sets generated by \thething{}, ML engineers could not continue to iteratively tune parameters of their algorithms to come up with a precise malware detection solution.

        \begin{table}[h]
            \footnotesize
                \begin{center}
                \caption{Confusion matrix of the random forest classifier.}
                \begin{tabular}{|c|c|c|}
                    \hline
                    -              &  act. benign   &   act. malware  \\ \hline
                    pred. benign   &  9987     &   2091    \\ \hline
                    pred. malware  &  13       &   8001   \\ \hline 
                \end{tabular}

                \end{center}
        \end{table}

        \begin{table}[h]
            \footnotesize
                \begin{center}
                \caption{Precision, recall and accuracy metrics for the malware class.}
                \begin{tabular}{|c|c|}
                    \hline
                    Class & Malware \\ \hline
                    Precision &  0.9983  \\ \hline
                    Recall    &  0.7928  \\ \hline 
                    Accuracy  &  0.8952  \\ \hline
                \end{tabular}
                \end{center}
        \end{table}

        \begin{table}[h]
        	\footnotesize
        	\begin{center}
        		\begin{tabular}{|c|c|}
        			\hline
					Class & Benign \\ \hline
					Precision &  0.8268  \\ \hline
					Recall    &  0.9987  \\ \hline 
					Accuracy  &  0.8952  \\ \hline
        		\end{tabular}
        		
        		\caption{Precision, recall and accuracy metrics for the benign class.}
        	\end{center}
        \end{table}

        For further verification of the ML algorithms under development, the \thething{} setup can be switched from collecting mode to enforcing mode. Rather than artificially creating traffic, the framework will then operate as an ICAP server classifying traffic using the ML algorithm under development. This allows to immediately evaluate malware detection solution against real traffic, while they are developed.
      
    \subsection{Changes of HTML/JavaScript Malware over Time}
    \label{subsec_evo_mal}
\begin{figure}[tb]
			\centering
            \includegraphics[width=\columnwidth]{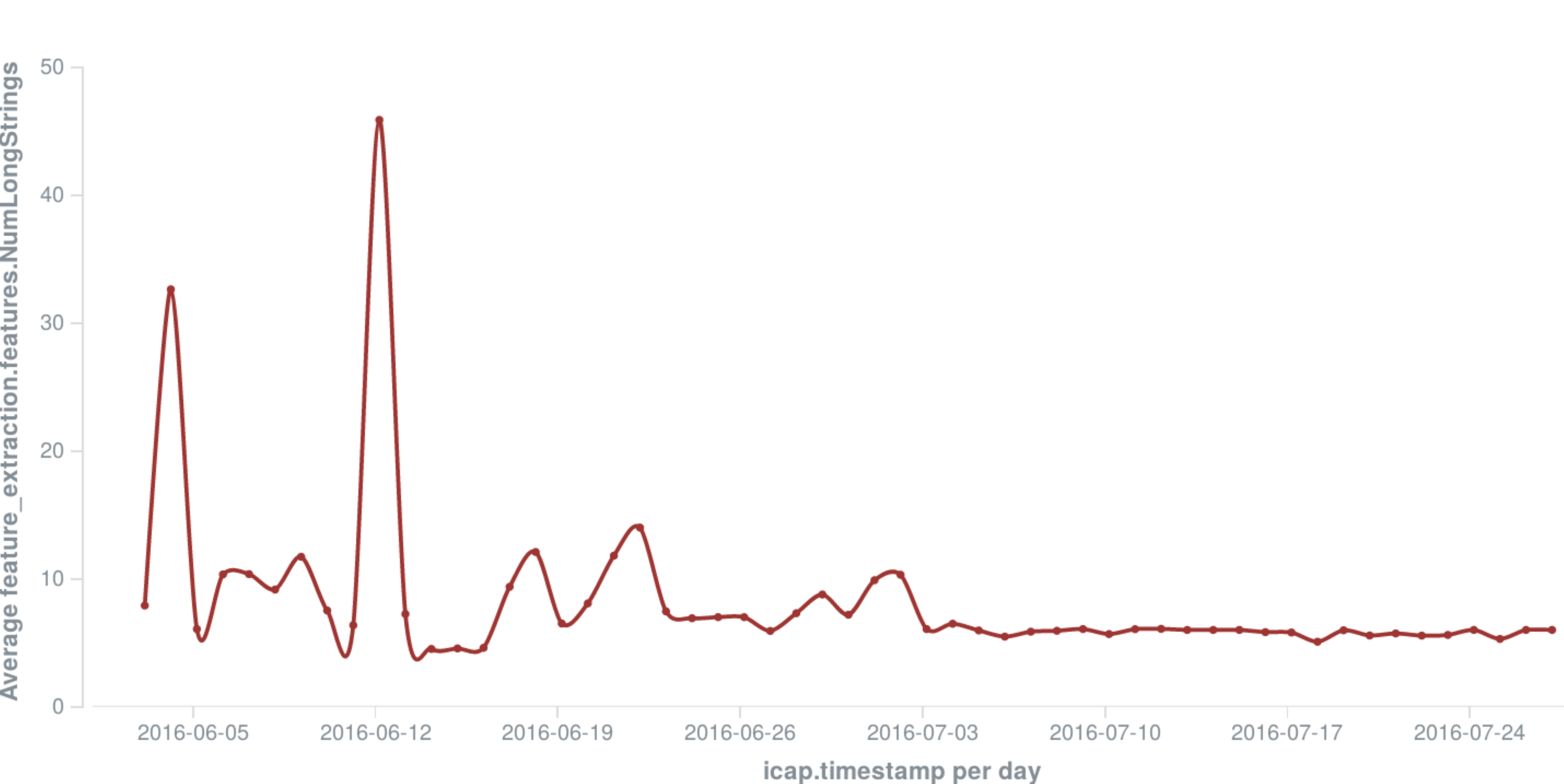}
            \caption{Average count of long strings within malicious samples}
            \label{fig:collection_progress}
          \end{figure}

      Extracting features from HTML and JavaScript documents allows us to make certain statements about malware based on these technologies. 
      For example, with the extracted HTML/JS features, it is possible to recognize the trends in circulating malware.
      Aligned with the collection progress~\ref{fig:collection_progress}, the most common features are (in terms of presence in the collected malware samples): 

      \paragraph{Long Strings}
        A common characteristic among JavaScript-Malware is the presence of long strings (>= 40 characters). 
        Around 95.05 percent of the malicious samples contain at least one long string.
        Whereas the benign sample set features long strings in 54.91 percent of the samples.

      \paragraph{Form Strings}
        Another common theme in HTML-/JavaScript-Malware is the presence of \texttt{form} strings within HTML-/JavaScript documents.
        Of all malware samples, 94.33 percent used at least one \texttt{form} string. 
        Benign JavaScript and HTML contained \texttt{form} strings in 33.50 percent of all samples.
        \\
		\begin{figure}[tb]
			\centering
            \includegraphics[width=\columnwidth]{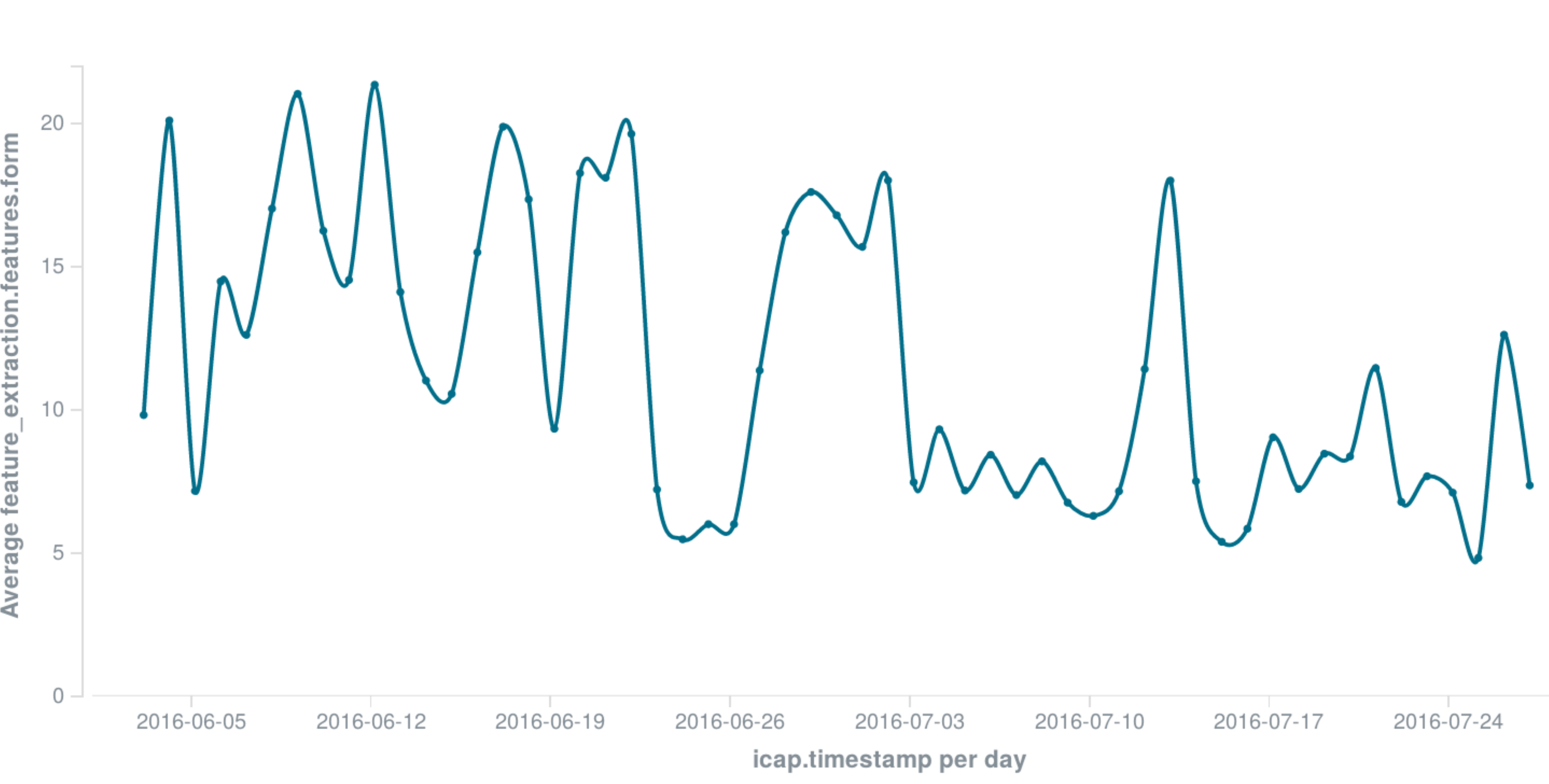}
            \caption{Average count of form strings within malicious samples.}
      \end{figure}
       \\

      \paragraph{Iframes}
        Iframes are also heavily used in the collected malware samples. 
        35.38 percent of the malicious samples contained Iframes.
        As compared to the benign set, where 18.56 percent contained Iframes.
        \\
		\begin{figure}[tb]
			\centering
            \includegraphics[width=\columnwidth]{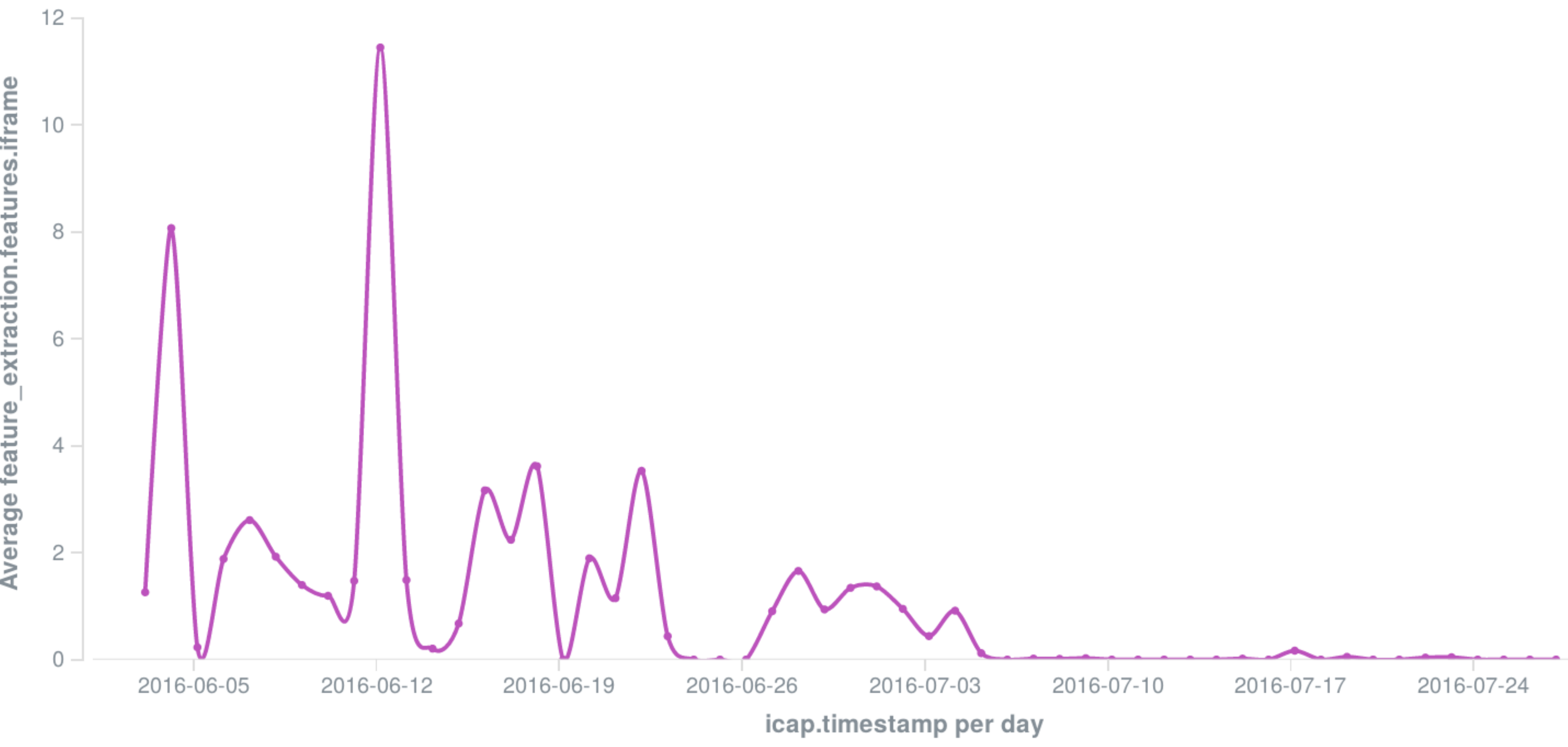}
            \caption{Average count of Iframes within malicious samples.}
      \end{figure}
      \\



\section{Conclusion}

In this report we motivated the need for the creation of large-scale data sets of web traffic which are precisely classified as malicious and benign, reflect the actual surfing behavior of users, and are free for research and product development to with, i.e. do not affect the privacy of any user. Our contribution to the research community is the design, implementation, and evaluation of \emph{\thething{}} that autonomously collects web traffic by simulating several users, classifies and extends data according to augmentation facilities, and extracts relevant features for machine learning training sets.

By a modular design and the use of standard interfaces such as ICAP, each functional part of the framework can be adopted to the specific needs of the user, such as the choice of different interaction strategies or malware classifiers. Further, by switching the framework from learning mode to enforcement mode, it can immediately serve as a testbed of malware detection solutions and thereby supports the test-development cycle.

Scalability of the system has been one of the main design criteria and was assessed in a two-months data collection period during which more than 43 million request/response pairs have been gathered and classified. The cloud-based deployment on an OpenStack/MongoDB/Elasticsearch infrastructure supports full scalability of user agents, data classification and storage and can thus be extended to any desired scale. Analytics visualizations are based on the Kibana tool and can thus be adopted to the user's needs as necessary. Attacks against the framework itself are limited by strict isolation of user agents. 

By applying our framework to train classic machine learning algorithms, we demonstrate how it supports the design of ML and deep learning approaches to detect malicious web traffic. The main challenges we came across in this research were the reliable labeling of sources to pre-classify visited websites, which has been solved by a combination of multiple classifiers, as well as the simulation of user behavior which has been implemented by different interaction strategies as part of the user agents.

As up to date, publicly available web data sets are scarce and severely limited, we consider \thething{} an important contribution to the field, as it allows researchers to create data sets of high quality and any desired size. As a starting point, we generated a 500 GB web traffic data set including a sufficiently large number of malicious samples to create ML training sets from it.

\section*{Acknowledgments}
This work has been funded by the German Federal Ministry of Education and Research (BMBF) through the project BDSec -- Big Data Security 01IS14009C.

\bibliographystyle{plain}
\bibliography{references}

\newpage
\onecolumn
\appendix
        \begin{table}[h]
            \footnotesize
  			\begin{center}
                \caption{Extracted Features}
                \begin{tabular}{|p{5cm}|p{8cm}|}
                    \hline
                    \textbf{Feature}        &  \textbf{Explanation}  \\ \hline
                    NumclearAttributes & Number of \texttt{clearAttributes} being called \\ \hline 
                    Filesize & Length of the file \\ \hline 
                    crypt &  Number of \texttt{crypt} string \\ \hline 
                    NumWords &  Number of words \\ \hline 
                    ishtml & Is the document a valid HTML? \\ \hline 
                    NumLongStrings & Number of long strings (>= 40 characters) \\ \hline 
                    TotalEntropy & Entropy of the file content \\ \hline 
                    NumReassignmentOfSpecialObject & Number of special objects being reassigned (e.g., \texttt{this}) \\ \hline 
                    onerror &  Number of registered \texttt{onerror} events \\ \hline 
                    isjs & Is the document a valid JS-File?\\ \hline 
                    NumActiveXObject & Number of \texttt{ActiveXObjects} \\ \hline 
                    MaxStringEntropy & Entropy of all javascript strings \\ \hline 
                    NumKeywords & Number of javascript keywords \\ \hline 
                    NumfireEvent & Number of \texttt{fireEvent} function calls \\ \hline 
                    NumreplaceNode & Number of \texttt{replaceNode} function calls \\ \hline 
                    NumBracketLookups & Number of bracket lookups (e.g., \texttt{this["eval"]} \\ \hline 
                    ShellcodeProbability & Probability of shellcode being present in the document \\ \hline 
                    AvgStringLength & Average length of strings in the document \\ \hline 
                    EntropyDensity & Density of the calculated entropy \\ \hline 
                    NumattachEvent & Number of events being attached to DOM elements \\ \hline 
                    containsjstags & Number of \texttt{<script>}-tags \\ \hline 
                    TotalStringEntropy & Calculated entropy of all strings \\ \hline 
                    onunload & Number of \texttt{onunload}-events being used \\ \hline 
                    script & Number of \texttt{script} strings \\ \hline 
                    NumHTMLNodes & Number of nodes in a HTML document \\ \hline 
                    MaxStrLen & Length of the longest string \\ \hline 
                    IP\_address & Number of IP-address like patterns in the document \\ \hline 
                    NumBracketCalls & Number of function calls with bracket notation \\ \hline 
                    NuminsertAdjacentElement & Number of insertAdjacentElement function calls \\ \hline 
                    NumNodes & Number of nodes in the Javascript-AST \\ \hline 
                    ishtmlwithjse4x & Is the document a valid HTML document with embedded JSE4X? \\ \hline 
                    NumStrings & Number of strings in the document \\ \hline 
                    evil & Number of \texttt{evil} strings \\ \hline 
                    NumiframeString & Number of \texttt{iframe} strings in the javascript AST \\ \hline 
                    NumaddEventListener & Number of \texttt{addEventListener} function calls \\ \hline 
                    NumsetInterval & Number of \texttt{setInterval} calls \\ \hline 
                    scriptTagDataURLCount & Number of javascripts loaded with DataURLs \\ \hline 
                    htmlEventCount & Number of all registered HTML events \\ \hline 
                    AvgLinesize & Average line size of the document \\ \hline 
                    shell & Number of \texttt{shell} strings \\ \hline 
                    NumPackerFunctions & Number of calls to a packer\\ \hline 
                    parsingerror & Does a parsing error exist?\\ \hline 
                    ishtmlwithjs & Is this a HTML with normal javascript? \\ \hline 
                    onload & Number of \texttt{onload} routines \\ \hline 
                    NumsetTimeout & Number of \texttt{setTimeout} calls \\ \hline 
                    TotalStringLength & Length of all strings combined \\ \hline 
                    embed & Number of \texttt{embed} strings \\ \hline 
                    Numeval & Number of \texttt{eval} strings \\ \hline 
                    object & Number of \texttt{object} strings \\ \hline 
                    frame & Number of \texttt{frame} strings \\ \hline 
                    spray & Number of \texttt{spray} strings \\ \hline 
                    NumLongVarOrFunNames & Number of variables or functions with long names \\ \hline 
                    iframe &  Number of \texttt{iframe} strings \\ \hline 
                    isjse4x & Is this a JSE4X document? \\ \hline 
                    NumdispatchEvent & Number of \texttt{dispatchEvent} function calls \\ \hline 
                    form & Number of \texttt{form} strings  \\ \hline 
                    NumFunctionCalls & Number of direct function calls \\ \hline 
                    onbeforeload & Number of \texttt{onbeforeload} events being used \\ \hline
                \end{tabular}
                \label{extracted_features}
        	\end{center}
        \end{table}

\end{document}